\documentclass[pra,
twocolumn,
superscriptaddress,showpacs,floatfix,nofootinbib]{revtex4}
\usepackage{epsfig,amsmath,amsfonts,amssymb}
\usepackage{graphicx}
\usepackage[utf8]{inputenc}
\usepackage{boxedminipage}
\usepackage{epsfig,psfrag}
\usepackage{amsmath}
\usepackage{mathbbol, amsfonts}
\usepackage{lscape}
\usepackage{fancybox}
\usepackage{amsfonts,amssymb}
\usepackage{color}
\usepackage{subfig}
\newcommand{\beq}{\begin{equation}}
\newcommand{\eeq}{\end{equation}}

\newcommand{\beqa}{\begin{eqnarray}}
\newcommand{\eeqa}{\end{eqnarray}}
\def\ra{\rangle}
\def\la{\langle}
\usepackage{wrapfig}

\begin{document}
\title{% Time-Optimal
 High-Speed Driving  of  a Two-Level System}
\author{Gerhard C. Hegerfeldt}
\affiliation{Institut für Theoretische Physik, Universit\"at G\"ottingen,
Friedrich-Hund-Platz 1, D-37077 G\"ottingen, Germany}

\begin{abstract}
 A remarkably simple result is found for the optimal protocol of drivings for  a general two-level Hamiltonian which transports a given initial state to a given final state in minimal time, under additional conditions on the drivings.
If one of the three possible drivings is unconstrained in strength the problem is analytically completely solvable. A surprise arises for a class of states when one driving  is bounded by a constant $c$ and the other drivings are constant. Then, for large $c$, the optimal driving is of type bang-off-bang and for increasing $c$ one recovers the unconstrained result. However, for smaller $c$ the optimal driving can suddenly switch to bang-bang type. It is also shown that for general states one may have a multistep protocol.
The present paper explicitly proves and considerably extends the author's results contained in  Phys. Rev. Lett. {\bf 111}, 260501 (2013).

\end{abstract}
\maketitle

\section{Introduction}

An important and challenging problem in many areas of physics is the fundamental task to drive a given initial quantum state to a prescribed target state in an optimal way by a `protocol', i.e. through a certain control of external fields and other parameters.   
These areas range from quantum computation \cite{Mo4}, fast population transfer in quantum optics \cite{Mup.4,Carmichael}, Bose-Einstein condensates \cite{Mup3}, nuclear magnetic resonance \cite{Levitt} to quite general atomic, molecular and chemical physics \cite{Rabitz, Rice}. In this context, what is meant by `optimal' depends on the particular question and situation \cite{Deffner}, and quite often one aims for a time-optimal protocol where the driving should reach the target state in the shortest possible time \cite{Caneva2009,Caneva2011,Caneva2013,Poggi2013,Garon2013,Liu2014,Anderson2013,Lloyd2014}. The connection of this minimal time with the so-called quantum speed limit time \cite{Caneva2009,Caneva2011,Ashhab2012,Caneva2013,Mukherjee,Deffner2013a,Deffner2013b,Campo,Taddei,Barnes,Xu} has been discussed in Ref. \cite{HePRL}. The adiabatic process, which is usually too slow,  can be modified by 
adding a so-called counterdiabatic term to achieve adiabatic dynamics with respect to the original Hamiltonian in a shorter time \cite{Berry1}, while `shortcuts to adiabaticity' (STA) \cite{Mup3,Chen} does not follow adiabatic states. For a  
comprehensive review of these and other approaches see Ref. \cite{Torr}. An experimentally important requirement for such protocols is fidelity. Small energy input as well as robustness may also play a role. Other approaches consider unitary time-development operators and aim to determine the optimal dynamics that leads from an initial $U(0)$ to a prescribed final propagator $U_F$ in minimal time \cite{Khaneja}.  Ref. \cite{Carlini} used a variational approach and Ref. \cite{Brody} a geometric approach to determine time optimal Hamiltonians under a trace condition and a condition on the separation of eigenvalues, respectively.

Considerable attention has focused in particular on two-level systems, or qubits, as the `simplest non-simple quantum problem' \cite{Berry2}. Ref.~\cite{Mo} studied the experimental implementation of control protocols  
where  two states  are coupled by a Landau-Zener  type Hamiltonian of the form $H= \Gamma(t)\sigma_3 + \omega \sigma_1$ 
where  $\sigma_i$ are the Pauli matrices and where in Ref. \cite{Mo} $\Gamma$ corresponds to quasi-momentum. 
A numerical analysis was performed in Ref. \cite{Caneva2009}. In Ref. \cite{HePRL} the optimal protocol was derived analytically and the minimal time to reach any given target state from any initial state was explicitly derived.
Another recent paper \cite{Gershoni} considered an analogous Hamiltonian and experimentally and numerically studied the time-optimal construction of single-qubit rotations under a strong driving field of finite amplitude.

In this paper we generalize our results of Ref. \cite{HePRL}  and provide explicit proofs. The system considered here is again a two-level system, but now governed by the completely general (traceless) Hamiltonian 
\beq \label{1.1}
H = \Gamma(t)\sigma_3 + \omega_1(t) \sigma_1 + \omega_2(t) \sigma_2. 
\eeq
The aim is to find  optimal drivings such that the time-development operator $U_H(t,0)$ associated with $H$ in Eq. (\ref{1.1}) evolves an initial state $|\psi_{\rm in}\ra$ at time $t=0$ to (a multiple of) a final state $|\psi_f\ra$ at time $T$ and to find the minimal time $T=T_{\rm min}$ for the following scenarios. 

Scenario (i): A single controllable driving field (`control') only, with the other drivings constant, in particular: (ia) no constraint on the driving field -- this is analytically completely solvable -- and (ib)  the driving field bounded by a constant. Scenario (ii): Two  time-dependent driving fields. If both controls are unconstrained, i.e. can be made arbitrarily large, the minimal time is zero \cite{Mo}. Therefore we consider: (iia) one control without constraint and the other constrained, and (iib) both constrained. For case (ib) we show explicitly that there is a transition from a bang - off - bang to a bang - bang protocol for a certain class of initial and final states, thus proving a result announced in Ref. \cite{HePRL}. It is also shown that in general one has a multistep protocol in this case.  The case of  three time-dependent controllable drivings is also investigated. If one of the controls is unconstrained  the problem is also analytically solvable.

In the following sections these scenarios are investigated. In Section \ref{Discussion} the physical relevance of the results is discussed, in particular with respect to finite switching time durations between different pulses.
As in Ref. \cite{HePRL} we use the Pontryagin maximum principle (PMP) \cite{PMP}  to derive the form of the optimal drivings, valid for both states and operators. Details are explained in Appendix~\ref{control}.

\section{Minimal driving time for a single control without constraint } \label{Gamma}

In case of a single controllable  driving it suffices to consider $\Gamma(t)$ as the control driving, with $\omega_1$ and $\omega_2$ fixed. The other cases are obtained by permuting the $\sigma_i$'s, as explained in more detail at the end of this section. Moreover, the case of two $\omega_i$'s can be reduced to the case $\omega_2=0$, as shown further below.
Thus we consider in this section the Hamiltonian
\beq \label{2.1}
H= \Gamma(t)\sigma_3 + \omega \sigma_1.
\eeq
The goal is to find an optimal driving $\Gamma(t)$ such that the corresponding time-development operator $U_H(t,0)$  evolves an initial state $|\psi_{\rm in}\ra$ at time $t=0$ to (a multiple of) a final state $|\psi_f\ra$ in minimal time $T=T_{\rm min}$, i.e.
\beq \label{2.4a}
U_H(T_{\rm min},0) |\psi_{\rm in}\ra = \lambda\, |\psi_f\ra~.
\eeq 
 If  $|\psi_{\rm in}\ra$ and $|\psi_f\ra$ are normalized to 1,  $\lambda$ is a phase factor, otherwise it also contains the ratio of the normalization factors. 

It is shown in Appendix~\ref{appendix1} that the optimal driving is given by  $\Gamma(t)\equiv 0$, except at the initial and final time \cite{HePRL}. When $\Gamma(t)\equiv 0$ the time-development operator becomes $U_H(t,0)=\exp\{-i\omega\sigma_1t\}$ which in general does not satisfy Eq.~(\ref{2.4a}). Therefore one needs initial and final $\delta$ -like pulses of zero time duration, e.g. a $\Gamma_{\rm in}^{(\epsilon)}(t)$ and $\Gamma_{\rm f}^{(\epsilon)}(t)$, $0\leq \epsilon$, with 
\beq \label{2.24}
\lim_{\epsilon \to 0} \int_0^\epsilon \Gamma_{\rm in,f}^{(\epsilon)}(t)dt \equiv \alpha_{\rm in,f}\,.
\eeq
In the initial and final pulse, $\omega$ drops out when, as required by Eq. (\ref{2.24}), $|\Gamma_{\rm in, f}| \to \infty$. The complete time-development operator for the optimal protocol from 0 to T is then of the form
\beq \label{U_H}
U_H(T,0) = e^{-i\alpha_{\rm f}\sigma_3}\, e^{-i\omega \sigma_1T}\, e^{-i\alpha_{\rm in}\sigma_3}.
\eeq
Since $\exp(i\pi\sigma_3) = -1$ one can independently add multiples of $\pi$ to $\alpha_{\rm in}$  and to $\alpha_{\rm f}$. 

For a given initial and final state one now has to determine all possible values of $\alpha_{\rm in,f}$ and $T$ such that Eq. (\ref{2.4a}) holds and then find the minimal $T$ among them.

We first consider the special case 
\beq \label{n2.1}
|\psi_{\rm in}\ra = \cos\frac{\theta_{\rm in}}{2} |0\ra + i \sin\frac{\theta_{\rm in}}{2}|1\ra
\eeq
or, equivalently, $|\psi_{\rm in}\ra =\begin{pmatrix} \cos\theta_{\rm in}/2 \\ i \sin\theta_{\rm in}/2 \end{pmatrix}$, with $0\leq \theta_{\rm in}\leq\pi$, and similarly for $|\psi_{\rm f}\ra$. From Eq. (\ref{2.4a}) one has
\begin{align} 
\lambda \begin{pmatrix} \cos\theta_{\rm f}/2 \nonumber\\ i \sin\theta_{\rm f}/2 \end{pmatrix} = \exp\{-i&\alpha_{\rm f}\sigma_3\} \exp\{-i\omega T\sigma_1\}\nonumber\\ &\times\exp\{-i\alpha_{\rm in}\sigma_3\}\begin{pmatrix} \cos\theta_{\rm in}/2 \\ i \sin\theta_{\rm in}/2 \end{pmatrix}\nonumber
\end{align}
\beq \label{n2.4}
=  \begin{pmatrix} e^{-i\alpha_+}\cos\omega T\cos\theta_{\rm in}/2 + e^{i\alpha_-}\sin\omega T\sin\theta_{\rm in}/2\\
i e^{i\alpha_+}\cos\omega T\sin\theta_{\rm in}/2 -i e^{-i\alpha_-}\sin\omega T\cos\theta_{\rm in}/2                    \end{pmatrix}
\eeq
where $\alpha_\pm \equiv \alpha_{\rm in} \pm \alpha_{\rm f} $.

We first note that Eq. (\ref{n2.4}) can, for example, be satisfied with $\alpha_{\rm in}=\alpha_{\rm f}=0$ and a (not necessarily minimal) $T = (\theta_{\rm in}-\theta_{\rm f})/2\omega$. For $\theta_{\rm in} > \theta_{\rm f}$ this is positive, but it is negative for $\theta_{\rm in} < \theta_{\rm f}$. Therefore, in the latter case we note that also $\alpha_{\rm in}=-\alpha_{\rm f} = \pi/2$ and $T = (\theta_{\rm f}-\theta_{\rm in})/2\omega$  satisfy Eq. (\ref{n2.4}).

We now claim that for the above special case the minimal time is given by $T_{\rm min} = |\theta_{\rm f}-\theta_{\rm in}|/2\omega $. To prove this we note that in order for Eq.  (\ref{n2.4}) to hold the ratios of the two components on each side have to be equal. Solving the resulting equation for $\tan \omega T$ one obtains by a straightforward calculation
\begin{align} \label{n2.5}
&|\tan\omega T|^2 =\nonumber\\
&\frac{\tan^2\theta_{\rm in}/2 +\tan^2\theta_{\rm f}/2 - 2\cos2\alpha_+\tan\theta_{\rm in}/2~\tan^2\theta_{\rm f}/2}{1+\tan^2\theta_{\rm in}/2\tan^2\theta_{\rm f}/2+ 2\cos2\alpha_-\tan\theta_{\rm in}/2~\tan^2\theta_{\rm f}/2}
\end{align}
If one could vary $\alpha_+$ and $\alpha_-$ independently (one can not do this because Eq. (\ref{n2.4}) has to hold!) this would become minimal for $\cos 2\alpha_+=\cos 2\alpha_- =1$ and the the right-hand side of Eq. (\ref{n2.5}) would become $|\tan(\theta_{\rm f}-\theta_{\rm in})/2|^2$.  But this actually coincides with the above example and therefore the claim is proved.

The case of general $|\psi_{\rm in}\ra$ and $|\psi_{\rm f}\ra$ will now be reduced to this special case. Any  $|\psi \ra$ can be written in the form
\beq \label{n2.6}
|\psi \ra = \tilde\lambda \begin{pmatrix} \cos\frac{\theta}{2} \\e^{ i\phi} \sin\frac{\theta}{2} \end{pmatrix} 
\eeq
with $0\leq \theta \leq \pi$, $0\leq \phi < 2\pi$ and with an irrelevant phase factor $\tilde\lambda$. We note that
\beq \label{n2.7}
 \begin{pmatrix} \cos\frac{\theta}{2}\\ e^{ i\phi} \sin\frac{\theta}{2} \end{pmatrix} =  e^{i(\phi-\pi/2)/2}e^{-i(\phi-\pi/2)\sigma_3/2}\begin{pmatrix} \cos\frac{\theta}{2}\\i \sin\frac{\theta}{2} \end{pmatrix}
\eeq
Hence Eq. (\ref{2.4a}) becomes
\begin{align} \label{n2.8}
\lambda \begin{pmatrix} \cos\theta_{\rm f}/2 \\ i \sin\theta_{\rm f}/2 \end{pmatrix}& =\exp\{i(\phi_{\rm f}-\pi/2)\sigma_3/2\}\nonumber\\ &\exp\{-i\alpha_{\rm f}\sigma_3\} \exp\{-i\omega T\sigma_1\}\exp\{-i\alpha_{\rm in}\sigma_3\}\nonumber\\&\exp\{-i(\phi_{\rm in}-\pi/2)\sigma_3/2\}\begin{pmatrix} \cos\theta_{\rm in}/2 \\ i \sin\theta_{\rm in}/2 \end{pmatrix}
\end{align}
Written in this form it becomes the previous case, and hence the optimal solution for the general case is 
\beq \label{n2.9}
T_{\rm min}= |\theta_{\rm f} - \theta_{\rm in}|/2\omega
\eeq
and
\begin{align} \label{n2.10}
\alpha_{\rm in} = \pi/4 - \phi_{\rm in}/2, ~~\alpha_{\rm f} = -\pi/4 + \phi_{\rm f}/2       ~~~\text {for~~} \theta_{\rm in}> \theta_{\rm f}\nonumber\\
\alpha_{\rm in} = -\pi/4 - \phi_{\rm in}/2, ~~\alpha_{\rm f} = \pi/4 + \phi_{\rm f}/2       ~~~\text {for~~} \theta_{\rm in}< \theta_{\rm f}
\end{align}
We recall that $\alpha_{\rm in}$ and $\alpha_{\rm f}$ are unique only up to a  multiple of $\pi$. 

The result in Eq. (\ref{n2.9}) was announced in Ref. \cite{HePRL} in a slightly different form. Writing $|\psi_{\rm in}\ra = i_0|0\ra + i_1|1\ra$ and
$|\psi_{\rm f}\ra = f_0|0\ra + f_1|1\ra$, Eq. (\ref{n2.9}) can be expressed as 
\beq \label{n2.11}
\cos \omega T_{\rm min} = |f_0i_0| + |f_1i_1|
\eeq
which is Eq. (22) of Ref. \cite{HePRL}.
\vspace{.5cm}

{\em Extension to Density Matrices.} The results are easily carried over to density matrices. Since a unitary transformation does not change the eigenvalues, $\rho_{\rm in}$ can only be transformed into $\rho_{\rm f}$ if both have the same eigenvalues, $\lambda_1$ and $\lambda_2$, say. The eigenvectors of any $\rho$ with $\lambda_1 \neq \lambda_2$ are orthogonal. The unitary time-development, $U_1$ say, that moves  $|\lambda_1\ra_{\rm in}$ in the shortest time  to $|\lambda_1\ra_{\rm f}$, up to a phase, also moves $|\lambda_2\ra_{\rm in}$ to $|\lambda_2\ra_{\rm f}$ in shortest time, up to a phase. Indeed, one has $U_1|\lambda_2\ra_{\rm in}\sim |\lambda_2\ra_{\rm f}$, by orthogonality. And if the time were not the shortest, one could choose a $U_2$ with shorter time. But this would also move $|\lambda_1\ra_{\rm in}$ to $|\lambda_1\ra_{\rm f}$ in shorter time, a contradiction. Thus the optimal time-development operator which moves  $\rho_{\rm in}$ to a $\rho_{\rm f}$ with the same eigenvalues is the one which moves an initial eigenvector to a final eigenvector in minimal time, which by Eqs.~(\ref{n2.9}) and (\ref{n2.10}) is given by $T_{\rm min}= |\theta_{\rm f}^{(1)} - \theta_{\rm in}^{(1)}|/2\omega =|\theta_{\rm f}^{(2)} - \theta_{\rm in}^{(2)}|/2\omega$.

\begin{figure}[tb]
\begin{center}
\includegraphics[width=.51\textwidth]{%hegerf/adiabaticity/
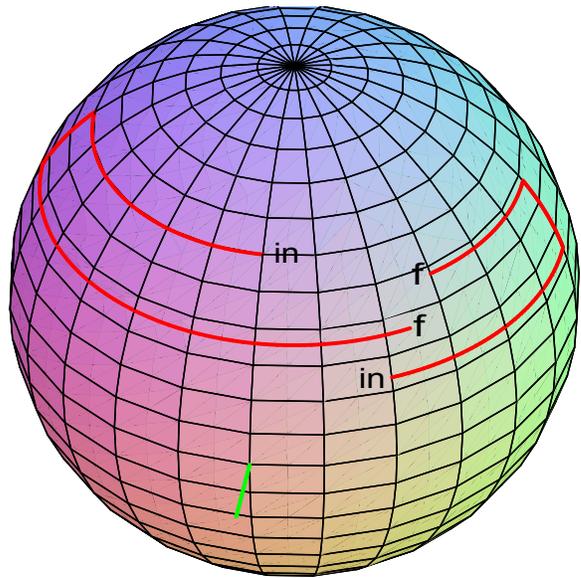}
\caption{Optimal protocol visualized on the Bloch sphere for unconstrained $\Gamma$: For $\theta_{\rm in}>\theta_{\rm f} $ the initial point is moved by a rotation around the $z$ axis in zero time to $\phi = \pi/2$ longitude, then rotated around the $x$ axis to the final latitude in time $T_{\rm min}=  |\theta_{\rm f} - \theta_{\rm in}|/2\omega$ and then rotated around the $z$ axis in zero time to the final position. For $\theta_{\rm in}<\theta_{\rm f}$ the initial point is first rotated around the $z$ axis in zero time to $\phi = -\pi/2$ longitude.  }
\label{unconstrained}
\end{center}
\end{figure}

\vspace{.5cm}
{\em Geometrical Considerations.} The physical state represented by $|\psi\ra$ of Eq. (\ref{n2.6}), i.e. $|\psi\ra \la\psi|$,  can be represented by the point $(\theta, \phi)$ on the unit sphere (`Bloch sphere') where $0\leq \theta \leq \pi$ is the polar angle and $0\leq \phi < 2\pi$ the azimuth angle. The operator $e^{-i\sigma_j \phi/2}$ corresponds to a rotation around the $j$ axis by the angle $\phi$. For $\theta_{\rm in}>\theta_{\rm f}$, the optimal time-development operator, which is determined by Eqs. (\ref{U_H}),  (\ref{n2.9}) and (\ref{n2.10}), corresponds, on the Bloch sphere, to a sequence of rotations, namely first a rotation of the point $(\theta_{\rm in}, \phi_{\rm in})$ around the $z$ axis to $(\theta_{\rm in}, \phi=\pi/2)$, then a rotation around the $x$ axis to $(\theta_{\rm f}, \phi =\pi/2)$, and finally a rotation around the $z$ axis to the final point $(\theta_{\rm f},\phi_{\rm f})$. If $\theta_{\rm in}<\theta_{\rm f}$, the first rotation is  around the $z$ axis to $(\theta_{\rm in}, \phi=-\pi/2)$. This is depicted in Fig. \ref{unconstrained}. The direct  path from `in' to `f', although geometrically the shortest, is here not the time-optimal path.

In particular, if the initial point lies in the $yz$ plane, e.g. given by $(\theta_{\rm in},  \pi/2)$, and if $\theta_{\rm in}>\theta_{\rm f}$, there is just a rotation around the $x$ axis by an angle $\theta_{\rm in} - \theta_{\rm f}$. This might seem obvious, but a formal proof -- either in spin space or on the Bloch sphere -- requires some effort. This is mirrored by the fact that for $\theta_{\rm in}<\theta_{\rm f}$ one first has to go to $ \phi = -\pi/2$ before rotating around the $x$ axis because otherwise the (positive) rotation angle would be given by $2\pi- (\theta_{\rm in} - \theta_{\rm f})$.

{\em Remark:} The results carry over in an analogous way if $\omega_1$ or $\omega_2$ is the control. E.g., if $\omega_1(t)$ is the control while $\Gamma$ is fixed and $\omega_2=0$ then one can put $\sigma_3' = \sigma_1$, $\sigma_1' = \sigma_3$,  $\sigma_2' = -\sigma_2$, $|0'\ra = (|0\ra + |1\ra)/\sqrt(2)$, and $|1'\ra = (|0\ra - |1\ra)/\sqrt(2)$. The Hamiltonian can then be written as $H= \omega \sigma_3' +\Gamma\sigma_1'$,  $|\psi_{\rm in}\ra = \cos\theta'_{\rm in}/2|0'\ra +\exp(i\phi'_{\rm in})\sin\theta'_{\rm in}/2|1'\ra$ and similarly for $|\psi_{\rm f}\ra$. Then everything carries over as before, only with `dashes', e.g.  Eq. (\ref{n2.9}) becomes $T_{\rm min}= |\theta'_{\rm f} - \theta'_{\rm in}|/2\Gamma$. For the visualization on the Bloch sphere the north pole now lies on the $x$ axis.

\section{Optimal fast driving under constraint} \label{1constraint}

We again consider the Hamiltonian of Eq. (\ref{2.1}) with the single control $\Gamma(t)$.
As is physically reasonable, it is now assumed that $\Gamma$ can not become arbitrarily large, i.e. 
\beq
|\Gamma(t)|  \leq c~.
\eeq
In Appendix \ref{constrained} it is shown that in this case the optimal driving will consist of intermittent periods with $\Gamma=\pm c$ and $\Gamma =0$. It will be shown in the following that the sequence, duration and number of these periods will depend on the bound $c$ and on the states involved. 

First we consider as initial and final state
\beqa \label{n3.1}
|\psi_{\rm in}\ra &=& \begin{pmatrix}\cos\theta_{\rm in}/2\\\sin\theta_{\rm in}/2 \end{pmatrix},~~~~~|\psi_{\rm f}\ra = \begin{pmatrix}\sin\theta_{\rm f}/2\\\cos \theta_{\rm f}/2 \end{pmatrix}\nonumber\\
\theta_{\rm in} &=& \pi/2+\alpha ,~~~~ \theta_{\rm f} = \pi/2-\alpha,  ~~~~\alpha > 0. \label{n3.1}
\eeqa
The states considered in Ref. \cite{HePRL} before Eq. (23) are in a different notation and are given by interchanging $|\psi_{\rm in}\ra$ and $|\psi_{\rm f}\ra$ and changing $\sin$ to $-\sin$ in Eq. (\ref{n3.1}). When $c \to \infty$ one should recover the result of the last section for the optimal time-development operator, and therefore we investigate an ansatz  where the initial and final $\delta$ pulse is replaced by a time development with $\Gamma=c$ and $\Gamma=-c$, respectively, and as yet unknown duration. One therefore arrives at
\beq \label{n3.2}
\lambda|\psi_{\rm f}\ra = e^{-i(- c\sigma_3+\omega \sigma_1) T_{-c}} e^{-i\omega\sigma_1 T_{\rm off}} e^{-i ( c\sigma_3+\omega \sigma_1) T_c}|\psi_{\rm in}\ra  
\eeq
where $T= T_{-c}+T_{\rm off}+T_c$  has to be minimized under the conditions $T_{\pm c}\geq 0$, $T_{\rm off}\geq 0$. This relation implies, as shown in  Appendix~\ref{with}, that $T_{-c}=T_c$ and that $T_{\rm off}$ can be expressed as a function of $T_c$ so that the total time $T$ becomes a function of $T_c$, $T=T(T_c)$. This latter function has to be minimized. Here we summarize the results and refer for the detailed derivation to Appendix~\ref{with}.

It turns out that for $c > \omega/\tan \alpha$, where $\alpha$ is defined in Eq. (\ref{n3.1}),  the optimal protocol is of bang-off-bang type, while for $c \leq \omega/\tan \alpha$ it is bang-bang. Explicitly, one has for $c \geq \omega/\tan \alpha $
\begin{align}
\!\!\!\!T_{\rm min}(c)&= 2T_c + T_{\rm off}\label{3.5b}\\  
T_c& = \frac{1}{\sqrt{c^2+\omega^2}}\arcsin\sqrt{\frac{c^2+\omega^2}{2c (c+\omega \tan \alpha)}}\label{3.5c}\\
T_{\rm off}&= \frac{1}{\omega}\arctan\frac{c \tan \alpha-\omega}{\sqrt{c^2+2c\omega \tan \alpha-\omega^2}}\label{3.5d}
\end{align}
For $c \to \infty$ one recovers the expression of the unconstrained case in Eq. (\ref{n2.9}). Furthermore,  as  $c \to \infty$, $T_c \to 0$ and $(c^2+\omega^2)^{1/2} T_{\rm c} \to \pi/4 $ so that the initial and final periods approach a $\delta$ pulse in $ \sigma_3$ of strength $\pm\pi/4$, as in the unconstrained case.  For $ c \leq \omega/\tan \alpha $ one has 
\begin{align}
T_{\rm min}(c) &=2T_c= \frac{2}{\sqrt{c^2+\omega^2}}\arcsin\sqrt{\frac{ \tan \alpha(c^2+\omega^2)}{2\omega(c+\omega \tan \alpha)}}\\
T_{\rm off}&= 0 ~.\label{3.5a}
\end{align}

For the optimal protocol, $\exp\{-i(c\sigma_3 +\omega\sigma_1) T_c\} $ transforms  the initial state $|\psi_{\rm in}\ra$ of Eq.~(\ref{n3.1}) into a state of the form
\begin{align}
&\lambda 
\begin{pmatrix}
\cos \theta'/2 \\ i\sin\theta'/2
\end{pmatrix} ~~~~~~\text{for $T_{\rm off} \neq 0$}\label{n3.3} \\
&\lambda 
\begin{pmatrix}
\cos\pi/4 \\ e^{i\phi}\sin\pi/4
\end{pmatrix} ~~~~ \text{for $T_{\rm off} =  0$}\label{n3.4}
\end{align}
as shown by a straightforward calculation.

\begin{figure}[tb]
\begin{center}
\includegraphics[width=.5\textwidth]{%hegerf/adiabaticity/
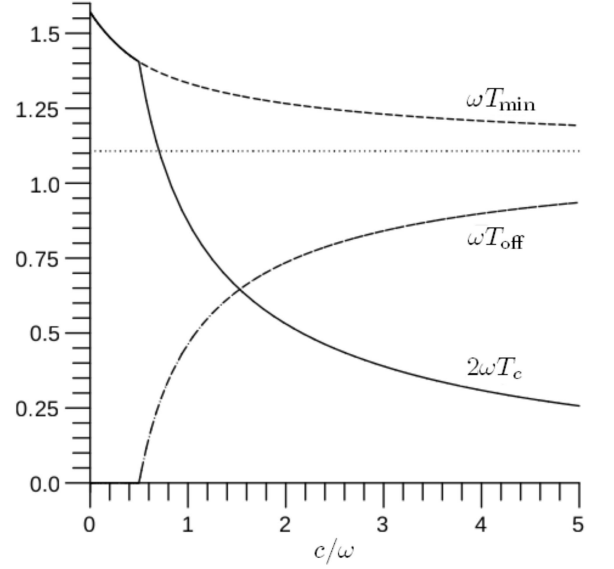}
\caption{$\omega T_{\rm min}$,  $\omega T_{\rm off}$ and $2\omega  T_c$, the double of the corresponding bang duration $\omega T_c$, as a function of $c/\omega$
 for  $\tan \alpha = 2$. For $c\to \infty$ one sees that $\omega T_{\rm min}$ approaches the unconstrained value $ \alpha$.
 For $c/\omega \leq 1/\tan\alpha = 0.5$ there is no period with $\Gamma(t)\equiv 0$, i.e $T_{\rm off} = 0$, so that for these values of $c$ the protocol is of bang-bang type. }
\label{Tmin}
\end{center}
\end{figure}

In Fig. \ref{Tmin},  $\omega T_{\rm min}$ is plotted as a function of $c/\omega$ for $\tan \alpha = 2$,  as well as the off duration $T_{\rm off}$, the asymptote $\alpha$ for the unconstrained case and $2 T_c$, the double of the corresponding individual bang duration.

\begin{figure}[tb]
\begin{center}
\includegraphics[width=.482\textwidth]{%hegerf/adiabaticity/
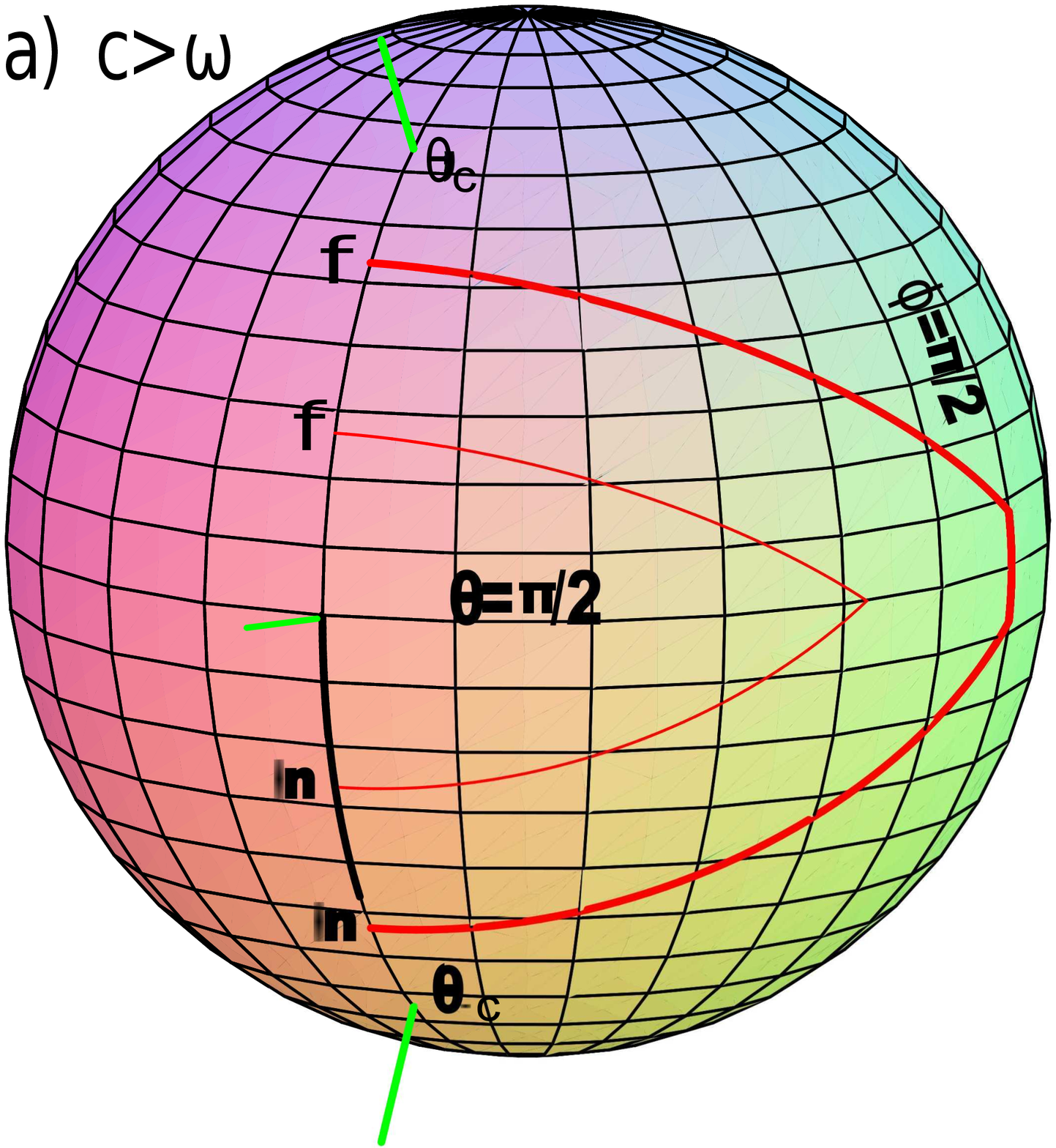}
\includegraphics[width=.5\textwidth]{%hegerf/adiabaticity/
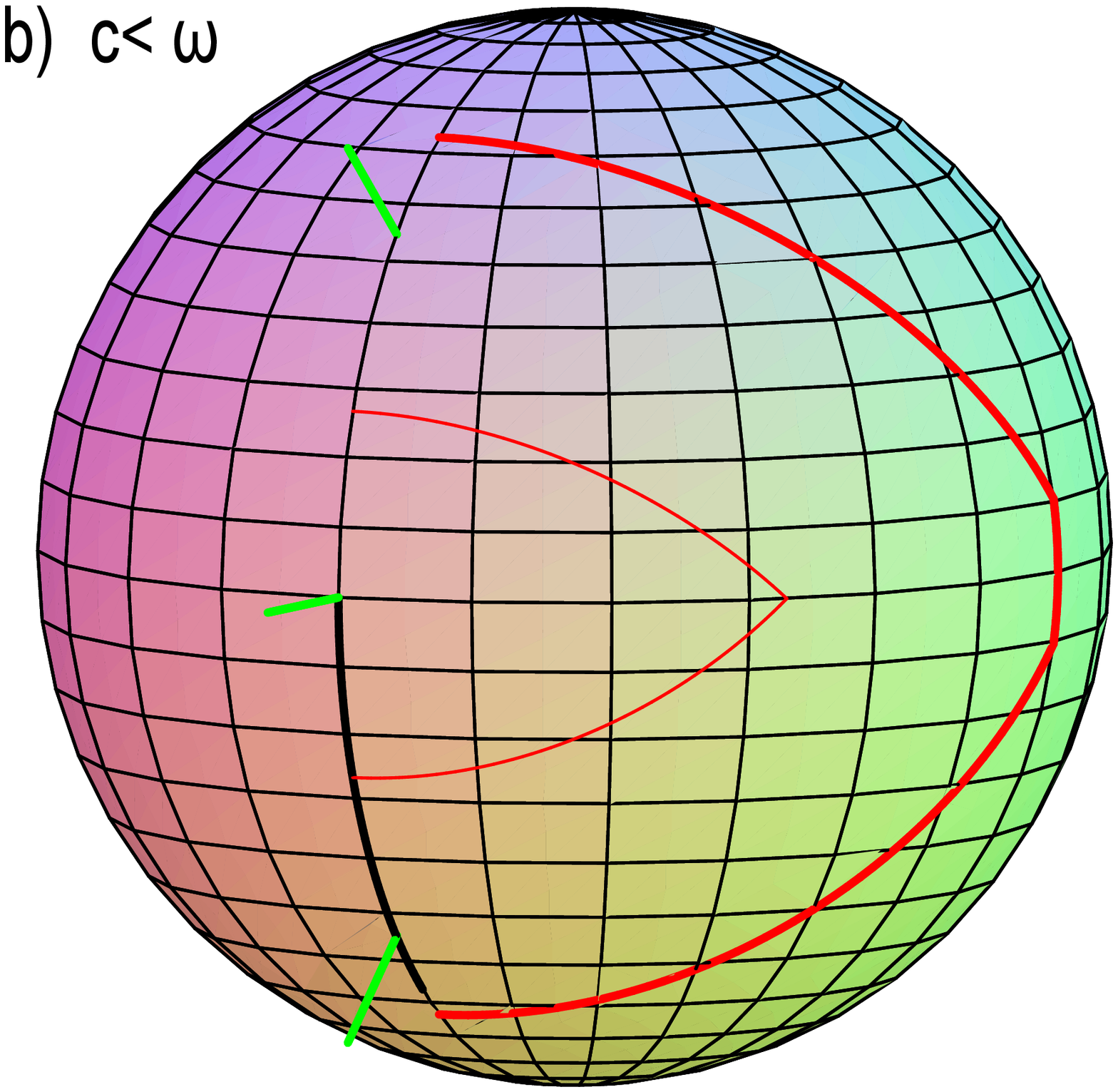}
\caption{Optimal protocol on the Bloch sphere for $\Gamma \leq c$ and 
initial and final state on longitude $\phi=0$ and symmetric with respect to the equator, with $\tan \theta_c = \omega/c$.  If $\theta_{\rm in} > \pi/2 + \theta_c$ (i.e. the initial point is outside the black line) the protocol is bang-off-bang, otherwise it is bang-bang. For $c> \omega$, $\theta_{-c}$ does not lie on the black line, while for $c < \omega$  it does.
For the bang-off-bang protocol there is first a positive rotation around the axis through $\theta_c$ until the 
longitude $\phi=\pi/2$ is reached. 
Then  a positive rotation around the $x$ axis rotates the point so far that a positive rotation around the $\theta_{-c}$ axis can bring it to the final destination. For the bang-bang protocol  the first  rotation moves the point to the equator at a longitude $\phi \leq \pi/2$ and then  a positive rotation around the axis through $\theta_{-c}$ leads to the final point. } 
\label{35}
\end{center}
\end{figure}

{\em Geometrical Considerations}. On the Bloch sphere, the optimal protocol for the initial and final state in Eq.~(\ref{n3.1})  has a very simple description. These  states  correspond to points on the Bloch sphere lying symmetrically with respect to the equator ($\theta=\pi/2$), with longitude $\phi = 0$ and polar angle $\theta_{\rm in} = \pi/2 + \alpha $ and $\theta_{\rm f}= \pi/2 - \alpha  $, respectively (cf. Fig. \ref{35}). The operator $\exp\{-i(c\sigma_3 +\omega\sigma_1) T_c\} $ corresponds to a rotation by the angle $2\sqrt{c^2+\omega^2} T_c $ around an axis with direction $(\omega,~0,~c)^t/\sqrt{c^2+\omega^2}$ which we call $\theta_c$ axis. On the Bloch sphere the $\theta_c$ axis goes through the point $\phi = 0$, $\theta=\theta_c$, with  $\sin\theta_c = \omega/\sqrt{c^2+\omega^2}$. The operator  $\exp\{-i(-c\sigma_3 +\omega\sigma_1) T_c\}$ corresponds to a rotation by the angle $2\sqrt{c^2+\omega^2} T_c $ around the  $\theta_{-c}$ axis  with direction $(\omega,~0,~- c)^t/\sqrt{c^2+\omega^2}$ which goes through the point  $\phi = 0$,  $\theta=\pi -\theta_c$.  The operator $\exp\{-i\sigma_1\omega T_{\rm off} \} $ corresponds to a rotation by the angle $2 \omega T_{\rm off}$ around the $x$ axis.

As a consequence of Eqs.~(\ref{3.5c}) - (\ref{n3.4}) the optimal protocol now proceeds as follows. The initial point is first rotated around the $\theta_c$ axis until it reaches the longitude $\phi = \pi/2$ or the equator, whichever is first. In the latter case, $T_{\rm off} = 0$ and one then continues directly with a rotation around  the $\theta_{-c}$ axis until, by symmetry, one reaches the final point. In the former case, one continues with a rotation around the $x$ axis (along the longitude $\phi = \pi/2$) until one reaches the circle around the $\theta_{-c}$ axis which goes through the final point and then continues along this circle to the final point. 

For given $c$, $\Gamma \leq c$,  initial points with $\pi/2 < \theta_{\rm in} \leq \theta_c +\pi/2$ (black line on longitude $\phi = 0$ in Fig.~\ref{35})  give rise to a bang-bang protocol ($T_{\rm off}  = 0$) while those with $\theta_{\rm in} > \theta_c +\pi/2$ have $T_{\rm off} > 0$. Note that for $c>\omega$ (i.e. $\theta_c < \pi/4$) the $\theta_{-c}$ axis goes through the black line while for $c<\omega$  it does not (cf. Fig. \ref{35} a) and b) ).

It should be noticed that  rotations around the $x$ axis and the $\theta_{\pm c}$ axis contribute in a different way to the total time since $ \sqrt{c^2 + \omega^2} > \omega$ and therefore a rotation by the same angle means more time for the rotation around the $x$ axis than for the $\theta_{\pm c}$ axis. Therefore there is a competition between the rotations and the optimal protocol is not a priori obvious.

\begin{figure}[tb]
\begin{center}
\includegraphics[width=.52\textwidth]{%hegerf/adiabaticity/
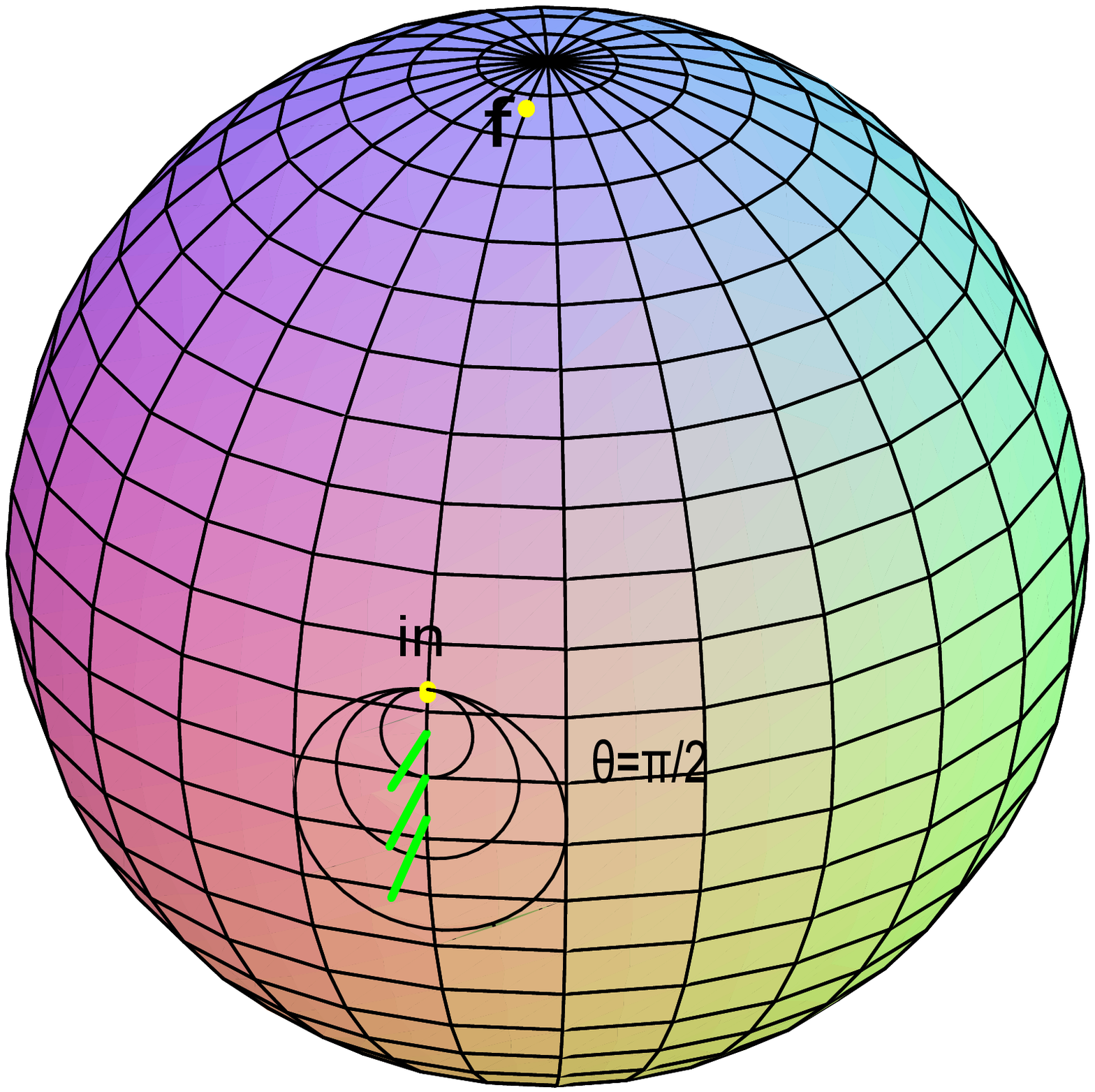}
\caption{Multistep protocols: For $c \ll \omega$, with $\theta_{\rm in}$ close to $\theta_c$ ($\tan\theta_c=\omega/c$) and $\theta_{\rm f}$ close to 0 more than three rotations are necessary to move the initial to the final point.}
\label{circles}
\end{center}
\end{figure}

For general initial and final states and  $\Gamma \leq c$ the situation becomes much more involved. A simple bang-off-bang protocol will in general have to be replaced by a multistep protocol. This is easily seen in the case of small $c$ where the initial point is on longitude $\phi=0$ and $\theta_{\rm in}$ is smaller than, but close, to $\theta_c$, while the initial point is close to the north pole  as in Fig. \ref{circles}. A first rotation around any of the  three axes will move the initial point further south, to a larger $\theta$, and it is apparent, that two further rotations will not be sufficient to move it to the final point since the available circle radii are too small. Hence one will need four or more rotations, and the optimization becomes much more complicated.


{\em Extension to Density Matrices.} As in the unconstrained case the results can be carried over to density matrices $\rho_{\rm in}$ and $\rho_{\rm f}$. To be connected by a unitary operator they must have the same eigenvalues $\lambda_1$ and $\lambda_2$. Then the time-optimal operator that moves $|\lambda_1\ra_{\rm in}$ to $|\lambda_1\ra_{\rm f}$  is also time-optimal for  moving $\rho_{\rm in}$ to $\rho_{\rm f}$.

\section{The general Hamiltonian}\label{reduction}

If both $\Gamma(t)$ and $\omega_1(t)$ are unconstrained the minimal time is zero \cite{Mo}. Therefore we first consider the case $\Gamma(t)$ unconstrained, $|\omega_1(t)|\leq \omega_{1\rm max}$ and $\omega_2 \equiv 0$. In Appendix  \ref{omega_1} it is shown that then for $\omega_1(t)$ the optimal choice is  either $\omega_1(t)= \omega_{1\rm max}$ or $\omega_1(t)= -\omega_{1\rm max}$. Moreover, there are no switches between $\pm \omega_{\rm max}$, and since the Hamiltonians with $\omega_1=\pm \omega_{1\rm max}$ are unitarily equivalent one can restrict oneself without loss of generality  to $\omega_1(t)= \omega_{1\rm max}$.  This is the case considered in Section \ref{Gamma}, and one thus has 
\beq \label{4.1}
T_{\rm min}= |\theta_{\rm f} - \theta_{\rm in}|/2\omega_{1\rm max},
\eeq
analogous to Eq.~(\ref{n2.9}).

We now consider the case where  one control is unconstrained and two controls are constrained. Without loss of generality we take
$\Gamma(t)$ as unconstrained, $|\omega_1(t)|\leq \omega_{1\rm max}$ and $\omega_2 \leq \omega_{2\rm max}$. In Appendix  \ref{2constraints} it is shown that then the minimal driving time is obtained by choosing
$\omega_i(t)\equiv\pm \omega_{i\rm max}$, $i= 1,~2$, and there are no switches between these values.  Moreover, the optimal time operator can be written in the form
\begin{align} \label{4.2}
U_H(T,0) =   e^{-i\alpha_{\rm f}\sigma_3}
 e^{-iT|\vec\omega_{\rm max}|\sigma_1 }  e^{-i\alpha_{\rm in}\sigma_3} 
\end{align}
where $|\vec\omega_{\rm max}|\equiv\sqrt{\omega_{1 \rm max}^2 +\omega_{2 \rm max}^2}$. The minimal time is given by 
\beq \label{4.3}
T_{\rm min}= |\theta_{\rm f} - \theta_{\rm in}|/2|\vec\omega_{\rm max}|,
\eeq
again analogous to Eq.~(\ref{n2.9}), and $\alpha_{\rm in}$ as well as $\alpha_{\rm f}$ are given by Eq.~(\ref{n2.10}).

When all three controls are constrained the result of Appendix \ref{3constrained} shows that the optimal controls are given by $\Gamma = 0,~ \pm \Gamma_{\rm max}$, $\omega_i=0,~ \pm \omega_{i {\rm max}}$ where in each combination at most one control vanishes. Hence the optimal time-development operator $U_(T,0)$ is an intermittent sequence of operators of the form
\begin{align}
&e^{-i(\pm \Gamma_{\rm max}\sigma_3 \pm \omega_{1 {\rm max}}\sigma_1
\pm \omega_{2 {\rm max}} \sigma_2 )T_j},~e^{-i( \pm \omega_{1 {\rm max}}\sigma_1 \pm \omega_{2 {\rm max}} \sigma_2 )T_j}\nonumber,\\
&e^{-i(\pm \Gamma_{\rm max}\sigma_3 \pm \omega_{2 {\rm max}} \sigma_2 )T_j},~~e^{-i(\pm \Gamma_{\rm max}\sigma_3 \pm \omega_{1 {\rm max}} \sigma_1 )T_j}
\end{align}
with  $\sum T_j= T $,  $T_j\ge 0$. For given initial and final state the ensuing minimization of $T$ will therefore in general lead to a multistep protocol, just as in Section \ref{1constraint}.

\section{Discussion}\label{Discussion}

We have investigated a quantum time-optimization problem  for a two-level system. For different scenarios it has been studied how to choose the three, possibly time-dependent, parameters (`controls') in the general Hamiltonian in such a way that the unitary time-development operator evolves a given initial state or density matrix to a given final state or density matrix in the shortest time possible. If two or more controls are unconstrained, i.e. if they can be made as large as one wants, the problem becomes trivial and  the minimal time is zero \cite{Mo}.

In this paper, for a single unconstrained control, both the optimal protocol and the associated time operator as well as the minimal time have been explicitly determined.    In case of a constrained control this has been carried through for a special class of initial and final states. It has also been shown that in the case of one unconstrained and two constrained controls the problem can be explicitly reduced to that of a single unconstrained control. A simple geometric interpretation on the Bloch sphere of the optimal protocol has been presented. 
For three constrained controls the general form of the optimal controls and of the unitary time-development has been determined.

If one of the controls, e.g. $\Gamma$, can experimentally be made much larger than the other drivings the situation becomes particularly simple. If $\Gamma_{\rm max} \gg \omega_1,~\omega_2$ say, then, to a good approximation, $\Gamma$  can be considered as unconstrained and the simple expressions from Sections \ref{Gamma} and \ref{reduction} apply in good approximation.

The results presented in this paper refer to an idealized situation, idealized insofar as instantaneous switching between different parameter values is experimentally not realizable but can only be approximated. However, the results provide a criterion for how close an experimentally realized protocol is to the ideal one. Moreover, as pointed out in Ref. \cite{HePRL}, if the time required for switching between different control values is small then the deviation from the ideal minimal time $T_{\rm min}$ is also small. 

Indeed, if there is an experimental switching time of duration $\epsilon>0$ to switch $\Gamma$ from $c$ to $0$ and  from 0 to $-c$, with $\omega\epsilon,\,c\epsilon \ll 1$, and if one retains $T_c$ and $T_{\rm off}$ from above, then  the
fidelity ${\cal F}$ can deviate from 1, but only slightly. More precisely, for the fidelity one has the bound 
 ${\cal F} > 1-2(\omega \epsilon + c\epsilon)$, instead of 1. The bound is 
 {\em independent} of the shape of the switching function. This can be shown
by first-order time-dependent perturbation theory. 

Moreover, instead of keeping $T_c$ and $T_{\rm off}$ from Eqs. (\ref{3.5c}) and (\ref{3.5d}) one can change them slightly in order
to increase the fidelity to 1, up to terms of second order in $\omega
\epsilon$ and $c \epsilon$. E.g., for a linear switching pulse one just
has to use $T_c - \epsilon/2$  and $T_{\rm off} - \epsilon$. For more general switching pulses a numerical approach seems to be needed.

If there are finite coherent times this implies an additional interaction. If the coherence times are much longer
than $T_{\rm min}$ this again implies only a small departure from ${\cal F}$=1, which can again be shown by perturbation theory. Therefore  coherence times much longer than $T_{\rm min}$  have only a small effect on $\cal F$. A quantitative investigation of this should be based on particular explicit models.

\begin{appendix}

\section{The Control Problem } \label{control}
To apply the Pontryagin maximum principle (PMP) \cite{PMP} we first parametrize the unitary time-development operator $U_H(t,0)$ in a convenient way. As a consequence of  the Eulerian  rotation angles  for the rotation group any $U\in {\rm SU(2)}$, in particular $U_H(t,0)$ for any traceless Hamiltonian such as in Eq. (\ref{1.1}),  can be written in the form 
\begin{align}
U_H(t,0) = \exp(-i\sigma_3\tau_3(t)/2)&\exp(-i\sigma_1\tau_1(t)/2)\nonumber\\
 &\times\exp(-i\sigma_3\tau_3'(t)/2) \label{2.5}
\end{align}
with three as yet unknown functions $\tau_3$, $\tau_1$ and $\tau_3'$.
We now differentiate both sides, equate the result with $\dot{U}_H =  -i\{\Gamma \sigma_3 +\omega_1 \sigma_1 +\omega_2 \sigma_2\}U_H$ and  multiply by 
$e^{i \sigma _3\tau_3/2}$ from the left and by $ ~ e^{i \sigma_3 \tau_3'/2}~e^{i \sigma_1 \tau_1/2} $ from the right. This gives
\beqa
&&\dot\tau_3 \sigma_3 + \dot\tau_1 \sigma_1 + \dot\tau_3' e^{-i \sigma_1 \tau_1/2}\sigma_3 e^{i \sigma_1 \tau_1/2 }\nonumber\\
&&=2 \Gamma\sigma_3 +2  e^{i \sigma _3\tau_3/2} \{ \omega_1 \sigma_1 +\omega_2 \sigma_2\} \ e^{-i \sigma_3 \tau_3/2 }.\nonumber
\eeqa
Using $e^{-i\sigma_3\tau_1/2}\sigma_1 e^{i\sigma_3\tau_1/2} =\cos\tau_1 \sigma_1 + \sin \tau_1\sigma_2$ etc.  one obtains
\begin{align}
&\dot\tau_3 \sigma_3 + \dot\tau_1 \sigma_1 + \dot\tau_3' (\cos\tau_1\,\sigma_3  -\sin\tau_1 \, \sigma_3)
= 2\Gamma\, \sigma_3 \nonumber\\
 &+2 \omega_1 (\cos\tau_3 \, \sigma_1 - \sin\tau_3 \, \sigma_2) + 2 \omega_2 (\cos\tau_3  \, \sigma_2 + \sin\tau_3 \, \sigma_1) .
\end{align}
Since the $\sigma_i$'s are linearly independent this leads to a system of three equations. With
\beq \label{2.8a}
 \vec\omega \equiv \begin{pmatrix}\omega_1\\ \omega_2 \end{pmatrix},~~
\vec e_r(\tau_3) \equiv \begin{pmatrix}\cos \tau_3\\ \sin \tau_3 \end{pmatrix},~~\vec e_\phi(\tau_3)\equiv \begin{pmatrix}-\sin \tau_3\\ \cos \tau_3 \end{pmatrix}
\eeq
they can be written as 
\beqa \label{2.9}
\dot \tau_1 &=& 2\vec\omega \cdot \vec e_r(\tau_3)\nonumber\\
\dot \tau_3' &=& - 2\vec\omega \cdot \vec e_\phi(\tau_3)/\sin \tau_1\\
\dot\tau_3  &=& 2\Gamma + 2\vec\omega \cdot\vec e_\phi(\tau_3) \, \cos\tau_1/\sin\tau_1~.\nonumber
\eeqa

The PMP deals with finding an optimal control function $u^*(t)$ (or possibly several control functions) such that a given cost function $J$ of the form $J=\int_0^{t_1} L(u(t),...)dt$, where $L$ is a function of $u(t)$ and some state functions and their derivatives, is minimized for $u(t)=u^*(t)$.
Here, the time $T$ required for the protocol is to be minimized, $J=T$, and since one can write $T=\int_0^T1\,dt$ one has $L\equiv 1$. 

We first  consider the case 
\beq \label{omega}
H = \Gamma(t)\sigma_3 + \omega\sigma_1
\eeq
and choose $u(t)=\Gamma(t)$  as the control and $\omega$ constant. The PMP then introduces the `control Hamiltonian'
\beq \label{2.12}
H_c = -L + p_1 \dot\tau_1 + p_3 \dot\tau_3 + p_3' \dot\tau_3'~,  ~~~~ L\equiv 1, 
\eeq
with as yet unknown functions $p_i(t)$ and where one inserts the $\tau_i$ derivatives from Eq. (\ref{2.9}), with $\Gamma$ replaced by $u$. Thus one obtains
\beqa \label{2.12a}
H_c = -1 &+& 2\omega p_1 \cos\tau_3 + 2\omega p_3'\sin \tau_3/\sin \tau_1 \nonumber\\
&+& 2 p_3( u -\omega \sin \tau_3\cos \tau_1/\sin \tau_1)~. 
\eeqa
 Then $H_c$ assumes its maximum for $u=u^*$, the optimal control, and in addition one has 
\beq \label{2.13}
\dot p_i = -\partial H_c/\partial \tau_i
\eeq
 when $u=u^*$, and similarly for $p_3'$.
Moreover, $H_c$ is constant along the optimal trajectory, and this constant is zero if the terminal time is free (i.e. not fixed), as in the present case.
In the following the asterisk on $u^*$ will be omitted. 

\subsection{Unconstrained $\Gamma$} \label{appendix1}

If $u$ is unrestricted, the maximality of $H_c$ gives $\partial H_c / \partial u = 0$, and by Eq. (\ref{2.12a}) this gives 
\beq \label{2.13a}
\frac{\partial H_c}{\partial u} = p_3=0 ~.
\eeq
This and Eq. (\ref{2.13}) give
\beqa 
\dot p_3&=& - \frac{\partial H_c}{\partial \tau_3}\nonumber\\&=& -2\omega p_1 \sin \tau_3 + 2\omega p_3' \cos \tau_3/\sin\tau_1=0 \label{2.15}\\
\dot p_1 &=& - \frac{\partial H_c}{\partial \tau_1} = 2\omega p_3' \sin\tau_3 \cos\tau_1/\sin ^2\tau_1\label{2.16}\\
\dot p_3' &=& - \frac{\partial H_c}{\partial \tau_3'} =  0, ~~~~p_3'= {\rm const}~\equiv~c_3'.
\eeqa
The relation $H_c = 0$ gives
\beq \label{2.18}
2\omega p_1 \cos\tau_3 + 2\omega c_3' \sin\tau_3/\sin\tau_1 = 1~.
\eeq
Multiplying this by $\sin\tau_3$ and Eq. (\ref{2.15}) by $\cos\tau_3$ and adding leads to
$$
\frac{2\omega c_3'}{\sin\tau_1}\sin^2\tau_3 + \frac{2\omega c_3'}{\sin\tau_1}\cos^2\tau_3 = \sin\tau_3
$$
and thus to
\beq \label{2.19}
2\omega c_3'/\sin\tau_1 = \sin\tau_3~.
\eeq
Insertion in Eq. (\ref{2.15}) gives
$$
(2\omega p_1 - \cos\tau_3)\sin\tau_3 = 0 
$$
and insertion in Eq. (\ref{2.18}) gives
$$
(2\omega p_1 - \cos\tau_3)\cos\tau_3= 0
$$
and therefore
\beq \label{2.19a}
2\omega p_1= \cos\tau_3,~~~ \dot p_1 = -\frac{\dot\tau_3}{2\omega}\sin\tau_3~.
\eeq
Inserting for $\dot\tau_3$ from Eq. (\ref{2.9}) leads to 
$$
 \dot p_1 = -\frac{1}{2\omega}(2\Gamma - 2\omega \sin\tau_3\, \cos\tau_1/\sin\tau_1)\sin\tau_3
$$
On the other hand, from Eqs. (\ref{2.15}) and (\ref{2.19}) one has
$$
\dot p_1 =  \sin^2\tau_3  \cos\tau_1/\sin\tau_1
$$
and these two equations imply
\beq \label{2.20}
\Gamma \sin\tau_3 = 0.
\eeq
Hence in any open interval in which $\Gamma \neq 0$ one has $\sin \tau_3 = 0$ and thus $\dot\tau_3 =0$, which also implies $\Gamma = 0$, by Eq. (\ref{2.9}). 
Hence in the unconstrained case the optimal choice for $\Gamma$ is $\Gamma(t) \equiv 0$, except possibly at the boundary points of the time interval. 

Note that so far the initial and final state have not come into play, and the result equally applies to operators.

\subsection{Constrained $\Gamma$}\label{constrained}

As is physically reasonable, it is now assumed that $\Gamma$ can not become arbitrarily large, i.e. 
\beq
|\Gamma(t)| \equiv |u(t)| \leq c~.
\eeq
In Eq. (\ref{2.12a}), the only term in $H_c$  which contains $u$ is of the form $2 p_3 u$, and hence for $H_c$ to become maximal one must have
\beq \label{3.2}
u(t) = 
\begin{cases}
~~\, c & \text{for $p_3(t)>0$}\\
-c & \text{for $p_3(t)<0$.}
\end{cases}
\eeq
If $p_3(t) \equiv 0$ in some time interval then  the argument from Eqs. (\ref{2.15}) - (\ref{2.20}) gives again $u(t)=\Gamma(t)=0$ in this interval. 
Hence the optimal driving will consist of intermittent periods with $\Gamma=\pm c$ and $\Gamma =0$. The sequence, duration and number of these periods will depend on $c$ and on the initial and final state. For $c \to 
\infty$ one should expect to recover the unconstrained case.

\subsection{Unconstrained $\Gamma$ and constrained $\omega_1(t)$} \label{omega_1}

If one allows in the Hamiltonian of Eq. (\ref{omega}) unconstrained $\Gamma(t)$ and $\omega(t)$ then the minimal time is 0 \cite{Mo}. We therefore consider here the case $\Gamma$ unconstrained and $|\omega(t)| \leq \omega_{\rm max}$. Then one can introduce two controls, $u = \Gamma$ and $u_1 = \omega$, which lie in the region $\{(u, u_1), -\infty <u <\infty,~-\omega_{\rm max}\leq u_1 \leq \omega_{\rm max} \}$. A maximum of $H_c$ can either lie in the interior of this region or on the boundary. 

The boundary of the region consists of the two straight lines $\{-\infty< u <\infty, u_1= -\omega_{\rm max}\}$ and $\{-\infty<u <\infty, u_1=\omega_{\rm max}\}$. A maximum on the boundary implies either $u_1=-\omega_{\rm max}$ or $u_1=\omega_{\rm max}$ and $\partial H_{\rm c}/\partial u = 0$. The latter gives $p_3=0$, as before. Moreover, the coefficient $p_1 \cos\tau_3 + p_3'\sin \tau_3/\sin \tau_1$ of $u_1 = \omega$ in Eq. (\ref{2.12a}) can not vanish because this would result in the contradiction $H_c = -1$. Hence there are no switches between $\pm \omega_{\rm max}$ and therefore $\omega(t)= \omega_{\rm max}$ or $\omega(t)= -\omega_{\rm max}$ throughout. Since $\sigma_3\sigma_{1,2}\sigma_3 = -\sigma_{1,2}$ the Hamiltonians with  $\omega(t)= \omega_{\rm max}$ and  $\omega(t)= -\omega_{\rm max}$ are unitarily equivalent and one can restrict oneself without loss of generality  to $\omega(t)= \omega_{\rm max}$.

If a maximum were in the interior this would imply that also $\partial H_{\rm c}/\partial u_1 = 0$ and this would lead to the contradiction $H_c = -1$, as before.

\subsection{Unconstrained $\Gamma$, constrained  $\omega_1(t)$ and $\omega_2(t)$} \label{2constraints}

Here we consider the Hamiltonian $H = \Gamma(t)\sigma_3 + \omega_1(t) \sigma_1 + \omega_2(t) \sigma_2 $ of Eq. (\ref{1.1}). If one allows more than one function to become unbounded then $T_{\rm min} \to 0$, by the symmetry $\sigma_i \to \sigma_j$. Therefore we consider the case $|\omega_i(t)|\leq \omega_{i,\rm max}$, $i=1,2$ and introduce the controls $u=\Gamma$ and $\vec u=\vec \omega$.

The control Hamiltonian is now given by
\begin{align} \label{A21}
H_{\rm c} = -1 + 2p_1 \vec u\cdot \vec e_r(\tau_3) - 2p_3' \vec u\cdot \vec e_\phi(\tau_3)/ \sin \tau_1\nonumber\\
+ 2p_3(u + \vec u\cdot \vec e_\phi(\tau_3) )\cos \tau_1/\sin \tau_1.
\end{align}

The region of allowed controls is the infinite slab $\{(u,u_1,u_2);-\infty < u < \infty, |u_1|\leq \omega_{1 \rm max}, |u_2|\leq \omega_{2 \rm max} \}$. Its boundary is given by the four sides of the slab. A maximum of $H_{\rm c}$ can either lie in the interior or on the boundary of the slab. 

{\em Case (i).} We first consider a maximum on one of the four corner lines. This will turn out to be the relevant case. Then $u_1 =\pm \omega_{1 \rm max}$, $u_2 =\pm \omega_{2 \rm max}$ and $\partial H_{\rm c}/\partial u = 0$. The latter gives $p_3=0$ and, because $\omega_{1,2}$ are now fixed, this implies similarly as before that $\Gamma =0$ in any open time interval. Again there are no switches, neither between $u_1=\omega_{1 \rm max}$ and $u_1=-\omega_{1 \rm max}$ nor between $u_2=\omega_{2 \rm max}$ and $u_2=-\omega_{2 \rm max}$. To prove this we write  $\vec\omega =(\pm\omega_{1 \rm max},\pm\omega_{2 \rm max})^t$
and $\vec\omega\cdot\vec\sigma = \omega_1\sigma_1 +\omega_2\sigma_2$. 
A possible optimal time operator could then be of the form
\begin{align}\label{A22}
\exp\{-i\alpha_{\rm f}\sigma_3\}&\exp\{-iT_1\vec\omega^{(1)}\cdot \vec\sigma \}\dots\nonumber\\& \dots\exp\{-iT_n\vec\omega^{(n)}\cdot \vec\sigma \}\exp\{-i\alpha_{\rm in}\sigma_3\}.
\end{align}
Using a rotation around the $z$ axis one has
\beq
\exp\{-i\vec\omega\cdot \vec\sigma T\} = e^{i \phi\sigma_3/2} \exp\{-i|\vec\omega|\sigma_1 T\} e^{-i \phi\sigma_3/2}
\eeq
where $\cos\phi = \omega_1/|\vec\omega|$ and $\sin\phi = \omega_2/|\vec\omega|$. Then Eq. (\ref{A22}) can be written as
\begin{align}\label{A24}
e^{-i\alpha_{\rm f}\sigma_3} e^{i\phi_1\sigma_3/2}&e^{-iT_1|\vec\omega_{\rm max}|\sigma_1 } e^{i(\phi_2-\phi_1)\sigma_3/2}\nonumber\\ &\dots e^{-iT_n|\vec\omega_{\rm max}| \sigma_1 } e^{-i\phi_n\sigma_3/2} e^{-i\alpha_{\rm in}\sigma_3}
\end{align}
where $|\vec \omega_{\rm max}| = \sqrt{\omega_{1 \rm max}^2 +\omega_{2 \rm max}^2}$. The total time $T= T_1 + \cdots + T_n$ has to be minimized. From Eq. (\ref{A24})  it follows that this is the same problem considered above for the case of unconstrained $\Gamma$, with $\omega$ replaced by $|\vec \omega_{\rm max}|$. As in that case,  for the minimal $T$ there is therefore only a single $T_i$ in Eq. (\ref{A22}) and one has $T=T_1$. Moreover, $T_{\rm min}$ is again given by Eq. (\ref{n2.10}), with $\omega$ replaced by $|\vec \omega_{\rm max}|$.

{\em Case (ii).} A maximum can also lie within a side of the of the slab boundary, e.g. on $\{(u,u_1,u_2);-\infty < u < \infty, u_1= \pm \omega_{1 \rm max} , |u_2|\leq \omega_{2 \rm max} \}$. This case can be ruled out as follows. One has $\partial H_{\rm c}/\partial u = 0$ and $\partial H_{\rm c}/\partial u_2 = 0$, while $u_1= \pm \omega_{1 \rm max}$. Similarly as in Eqs. (\ref{2.13}) - (\ref{2.19a}) one obtains from $H_c=0$ and $\dot p_3=0$
\beqa 
p_1 &=& \frac{1}{2|\vec u|^2}\vec u \cdot \vec e_r(\tau_3)\\
p_3'&=& - \frac{\sin \tau_1}{2|\vec u|^2}\vec u \cdot \vec e_\phi(\tau_3).
\eeqa
Inserting this into the equation resulting from $\partial H_{\rm c}/\partial u_2 = 0$ gives $u_2=0$. Since $\omega_1=\pm \omega_{1 \rm max} $ one arrives back at the situation considered in the first subsection of the appendix, with $\omega =\pm \omega_{1 \rm max} $, and therefore $\Gamma =0$. Hence, in the present case, there may be time intervals in which the time development is given by  $\exp\{\mp i\omega_{1 \rm max}T\sigma_1\}$. For the total time development this means that
in Eq. (\ref{A22}) some of the factors may be replaced by $\exp\{\mp i\omega_{1 \rm max}T_j\sigma_1\}$. But since $\omega_{1 \rm max} < \sqrt{\omega_{1 \rm max}^2 +\omega_{2 \rm max}^2}$ this would give a greater total time.

{\em Case (iii).} For a maximum in the interior of the slab one obtains again the contradiction $H_c = -1$.

Therefore only case (i) is realized. One has a fixed $u_{1,2}$ with
$u_i= \pm \omega_{i \rm max} $ (no switching). The optimal time operator is of the form
\begin{align} \label{A23}
U_H&(T,0) =
\nonumber\\ &  e^{-i\alpha_{\rm f}\sigma_3}
 e^{-i\phi\sigma_3/2}e^{-iT|\vec\omega_{\rm max}|\sigma_1 } e^{-i\phi\sigma_3/2} e^{-i\alpha_{\rm in}\sigma_3} 
\end{align}
where the $\phi$ terms can be absorbed in $\alpha_{\rm f}$ and $\alpha_{\rm in}$. The minimal time is given by Eq. (\ref{n2.9}), with $\omega$ replaced by $|\vec\omega_{\rm max}|=\sqrt{\omega_{1 \rm max}^2 +\omega_{2 \rm max}^2}$~.

\subsection{Three constrained controls} \label{3constrained}

When all three controls are constrained, i.e. $|u|=|\Gamma (t)|\leq \Gamma_{\rm max}$, $|u_1|=|\omega_1(t)|\leq \omega_{1\rm max}$ and $|u_2|=|\omega_2(t)| \leq \omega_{2\rm max}$, the situation becomes more complex. The region of allowed controls $(u,~u_1,~u_2)$ is now a finite rectangular box. The optimal controls could either lie in its interior or on one of its faces. The interior is ruled out as in case (iii) of the previous subsection. Neither can a maximum lie in the interior of one of the $(u,~u_1)$ faces, by the same argument as in case (ii) of the previous subsection, and the same holds for the interior of the $(u_1,~u_2)$ faces. The latter fact follows by making a unitary transformation with $\{\Eins -i\sum \sigma_j\}/2$ which transforms $\sigma_i$ into $\sigma_{i+1}$ cyclically and so interchanges the role of $\Gamma$, $\omega_1$ and $\omega_2$ cyclically. Hence a maximum can only lie in the interior of a line joining two corners of the box or on one of the corners. In the former case two controls are fixed and the derivative of $H_c$ with respect to the third vanishes. A similar argument as in Eqs. (\ref{2.15}) - (\ref{2.20}) yields the result that this control is zero.  

Hence the optimal Hamiltonian is an intermittent sequence of  Hamiltonians  with $u=0,~ \pm \Gamma_{\rm max}$, $u_i=0,~ \pm \omega_{i {\rm max}}$ where in each combination of controls at most one control can vanish.

\section{ Determination of $T_{\rm min}$ with constraint}\label{with}

In order to evaluate Eq. (\ref{n3.2}) we put 
\beqa \label{3.3}
\rho &\equiv& \sqrt{c^2 + \omega^2},~~ \tilde c \equiv c/\rho,~~ \tilde\omega = \omega/\rho\nonumber\\
\sigma_{\pm c} &\equiv& \pm\,\tilde c \,\sigma_3 + \tilde{\omega}\,\sigma_1. \eeqa
We note that $\sigma_{\pm c}^2 = 1$ and that $\sigma_1\sigma_{-c}\sigma_1 = \sigma_c$. Furthermore, one has  $|\psi_{\rm f}\ra =\sigma_1|\psi_{\rm in}\ra$.
As a consequence Eq. (\ref{n3.2}) can be written as
\beq \label{3.5b}
\sigma_1 e^{-i\sigma_c\rho T_{-c}} \sigma_1 e^{-i\omega \sigma_1T_{\rm off}} e^{-i \sigma_c \rho T_c}|\psi_{\rm in}\ra  = \lambda |\psi_{\rm in}\ra ,
\eeq
i.e. $|\psi_{\rm in}\ra $ is an eigenvector of the operator on the left-hand side. But then $|\psi_{\rm in}\ra $  is also an eigenvector of the trace-free part of the operator. Therefore, inserting
\beq
e^{-i\sigma_j \varphi} = \cos\varphi - i \sigma_j \sin\varphi~,~~~j=1,2,3,c,
\eeq
into Eq. (\ref{3.5b}) only those terms are relevant which are linear in $\sigma_{1,2,3,c}$ or are products  which can be reduced to such linear terms. The $\sigma_2$ term becomes
\begin{align}
\sigma_2 \{\tilde c \sin\rho T_{-c} &\cos \omega T_{\rm off} \cos\rho T_c\nonumber\\
 - \tilde c\cos&\rho T_{-c}\cos\omega T_{\rm off}\sin\rho T_c \}\nonumber\\
&=\sigma_2 \tilde c \sin\rho(T_c-T_{-c}) \cos\omega T_{\rm off}~.
\end{align}
Since $|\psi_{\rm in}\ra $  has only real components, since $\sigma_2$ has no real eigenvector and since the other terms will later turn out to be real, this term has to vanish. Hence we have $T_{-c} = T_c$.

The remaining $ \sigma_i$ terms are then calculated as 
\begin{align}
\sigma_1\{\cos^2\rho T_c\cos\omega T_{\rm off} +& (\tilde c^2-\tilde \omega^2) \sin^2\rho T_c\cos\omega T_{\rm off}\nonumber\\
 &-\tilde \omega \sin2\rho T_c \sin\omega T_{\rm off}
 \}\nonumber\\
-\sigma_3\{ \tilde c\sin2\rho T_c\sin&\omega T_{\rm off} + 2\tilde c\tilde\omega \sin^2\rho T_c \cos\omega T_{\rm off} \}\nonumber\\
\equiv \sigma_1\{I\} - \sigma_3&\{II\}.\label{3.8}
\end{align}
In order that $|\psi_{\rm in}\ra$ be an eigenvector of this operator,  the ratios of the first and second components of $\left\{\sigma_1\{I\} - \sigma_3\{II\}\right\}|\psi_{\rm in}\ra$  and of $|\psi_{\rm in}\ra$ have to be equal, and  this gives $\{I\}\cos\theta_{\rm in} + \{II\}\sin\theta_{\rm in} = 0$. Inserting from Eqs. (\ref{3.8}) and (\ref{n3.1}) a brief calculation gives, with $\alpha = \theta_{\rm in} - \pi/2$,
\begin{align} \label{3.11}
\tan \omega T_{\rm off} = \frac{\tan\alpha - 2\tilde\omega(\tilde c +\tilde\omega \tan \alpha)\sin^2\rho T_c}{(\tilde c + \tilde\omega \tan \alpha)\sin 2\rho T_c }\equiv \frac{N}{D}
\end{align}
which expresses $T_{\rm off}$ as a function of $T_c$. Now one has to minimize $T \equiv 2 T_c + T_{\rm off}(T_c)$  under the condition that $T_{\rm off}\geq 0$. It should be noted that one can restrict $T_c$ to $\rho T_c \leq \pi/2$, by Eq. (\ref{3.11}). Denoting the numerator in Eq. (\ref{3.11}) by $N$ and the denominator by $D$ one obtains 
\beq\label{B24}
T = 2T_c + \frac{1}{\omega}\arctan \frac{N}{D}
\eeq
One easily calculates $\partial N/\partial T_c = -2\omega D$ and using this one finds
\begin{align}
\frac{\partial T}{\partial T_c}& = 2 + \frac{1}{\omega} \frac{1}{1+N^2/D^2}\frac{D\,\partial N/\partial T_c - N\,\partial D/\partial T_c}{D^2}\nonumber\\
&= \frac{1}{\omega}\frac{1}{N^2+D^2}N\,(2\omega N -\partial D/\partial T_c)~.\label{3.13}
\end{align}
In order for $T$ to have an extremum, either $N$ or the term in brackets must be zero, i.e. there must be a time $T_c^{(1)}$ such that $N(T_c^{(1)}) = 0$
or there must be a time $T_c^{(2)}$ for which the term in brackets vanishes.
A brief calculation gives the conditions
\beq \label{3.15}
\sin^2(\rho T_c^{(1)}) = \frac{(c^2+\omega^2) \tan\alpha }{2\omega (c+\omega\tan \alpha)}
\eeq
\beq \label{3.16}
\sin^2(\rho T_c^{(2)}) = \frac{c^2+\omega^2}{2c(c+\omega\tan\alpha)}~.
\eeq
Since the sine function is bounded by 1 it follows from these expressions that 
 $T_c^{(1)}$ can exist only for 
\beq
 \frac{c}{\omega} \leq \frac{1}{\tan\alpha/2}
\eeq
while  $T_c^{(2)}$ exists only for 
\beq
\frac{c}{\omega} \geq \frac{1}{\cos\alpha} -\tan\alpha~.
\eeq

For $T_c=T_c^{(1)}$, i.e. $N=0$, one easily sees that $\partial^2T/\partial T_c^2$ is  positive if $\sin \rho T_c^{(1)} < 1/\sqrt{2}$. From Eq. (\ref{3.15}) it then follows that one has a minimum for $T_c=T_c^{(1)}$ if $c/\omega < 1/\tan\alpha$. Similarly, for $T_c=T_c^{(2)}$ the second derivative is positive if $T_c^{(2)} < T_c^{(1)}$ where one uses the fact that $N>0$ for $T_c < T_c^{(1)}$. Hence there is a minimum at $T_c=T_c^{(2)}$ if $T_c^{(2)} < T_c^{(1)}$. For $c/\omega = 1/\tan\alpha$ one has 
 $T_c^{(1)} = T_c^{(2)}$ and $ T_c^{(1)}= \pi/4\rho$. Moreover, $T_{\rm off}=0$ at this value of $T_c$ and $T_{\rm off}$ becomes negative for $T_c > T_c^{(1)} $. Hence one has to determine only the minimum  in the interval $0\leq T_c \leq  T_c^{(1)}$. Since 
it can not lie in the interior and since for $T_c \to 0$ one has $T\to \pi/2\omega$ while $T=2 T_c^{(1)} = \pi/2\rho$  at the other end point the minimum lies at $T_c= T_c^{(1)}$.

From these considerations it follows that for fixed $c$ the minimum of $T$ is obtained for $T_c = T_c^{(1)}$ if $c\leq \omega/\tan \alpha$, and for $T_c = T_c^{(2)}$ otherwise. Since $\tan\omega T_{\rm off} = 0$ for $T_c =T_c^{1}$ one has 
\beq
T_{\rm min} =
\begin{cases}
 2 T_c^{(1)} & \text{for $c \leq \omega/\tan \alpha$}\\
 2 T_c^{(2)} + T_{\rm off}(T_c^{(2)}) & \text{for $c \geq \omega/\tan \alpha $}~.
\end{cases}   
\eeq
From Eqs. (\ref{3.15}) and (\ref{3.16}) one then finally obtains Eqs. (\ref{3.5c}) - (\ref{3.5a}).

\end{appendix}

\end{document}